\documentclass[multphys,vecphys]{svmult}

\usepackage{makeidx}         
\usepackage{graphicx}        
\usepackage{multicol}        
\usepackage[bottom]{footmisc}

\makeindex             

\newcommand{\bea}{\begin{eqnarray}} 
\newcommand{\eea}{\end{eqnarray}}

\begin{document}

\title*{\bf Quantum Field Theory: Where We Are}
\author{{Klaus Fredenhagen}\inst{1}\and
{Karl-Henning Rehren}\inst{2}\and
{Erhard Seiler}\inst{3}}

\institute{II. Institut f\"ur Theoretische
Physik, Universit\"at Hamburg, \\  22761 Hamburg, Germany \\
\texttt{klaus.fredenhagen@desy.de}
\and Institut f\"ur Theoretische Physik, Universit\"at G\"ottingen,
\\ 37077 G\"ottingen, Germany \\
\texttt{rehren@theorie.physik.uni-goe.de}
\and Max-Planck-Institut f\"ur Physik, \\ 80805 M\"unchen, Germany \\
\texttt{ehs@mppmu.mpg.de}}

\maketitle

\vskip10mm
\begin{abstract}  We comment on the present status, the
  concepts and their limitations, and the successes and open problems 
  of the various approaches to a relativistic quantum theory of elementary 
  particles, with a hindsight to questions concerning quantum gravity and 
  string theory. 
\end{abstract}
\vskip36mm

\renewcommand{\thefootnote}{} 

\footnote{Contribution to: \it An Assessment of Current Paradigms in
  the Physics of Fundamental Phenomena, \rm to be published by
  Springer Verlag (2006).} 
\section{Introduction}

Quantum field theory aims at a synthesis of quantum physics with the 
principles of classical field theory, in particular the principle of
local\index{locality}ity. Its main realm is the theory of elementary
particles where it led to a far reaching understanding of the
structure of physics at subatomic scales with an often amazingly good
agreement between theoretical predictions and experiments.  
Typical observables in QFT are current densities or energy flow densities
which correspond to what is measured in particle physics detectors.
The original aim of QFT was to compute expectation values and correlation 
functions of the observables, and to derive scattering\index{scattering} 
cross section\index{cross section}s in high-energy physics. In the
course of development, QFT has widened its scope, notably towards the
inclusion of gravit\index{quantum gravity}ational interactions. 

Quantum Field Theory rests on two complementary pillars. The first 
is its broad arsenal of powerful modelling methods, both 
perturbat\index{perturbation theory}ive and
constructive\index{constructive QFT}. These methods are based on
quantiz\index{quantization}ation of classical interactions and the
gauge\index{gauge principle} principle, and have been tremendously
successful especially for the modelling of all the interactions of the
Standard Model of elementary particles. The perturbative treatment of
the Standard Model and its renormaliz\index{renormalization}ation, as
well as lattice\index{lattice approximation} approximations of Quantum
Chromodynamics (QCD\index{QCD}), give enormous confidence into the
basic correctness of our present  understanding of quantum
interactions. (For the impressive phenomenological support for the
Standard Model, we refer to Dosch's contribution to this volume.)
Despite these successes, however, establishing the Standard Model (or
part of it) as a mathematically complete and consistent quantum field
theory remains an unsettled challenge. 

The second pillar of QFT are axiomatic\index{axiomatic approach}
approaches, putting the theory on a firm conceptual ground, which have
been developped in order to understand the intrinsic features of a
consistent QFT, irrespective of its construction. In these approaches,
the focus is set on the fundamental physical principles which any QFT
should obey, and their axiomatic formulation in terms of the
observable features of a theory is addressed. 

In fact, several such axiomatic\index{axiomatic approach} approaches,
which have been shown to be partially but not completely equivalent,
are pursued. None of them indicates a {\em necessary} failure or
inconsistency of the framework of QFT. (Of course, this does not mean
that a realistic QFT should not include new Physics, say at the
Planck\index{Planck length} scale, cf.\ Sect.~\ref{gravity}.)

\section{Axiomatic Approaches to QFT}\label{axiom}

Axiomatic\index{Axiomatic approach} QFT relies on the fact that the
fundamental principles which every quantum field theoretical model
should satisfy are very restrictive. On the one hand this is a great
obstacle for the construction of models, on the other hand it allows
to derive a lot of structural properties  which a QFT necessarily
has. They often can be tested experimentally, and they provide a
criterion whether a construction of a model is acceptable. 

The main principles are: 
\begin{itemize}
\item the superposition\index{superposition principle} principle for
  quantum states, and the probabil\index{probabilistic interpretation}istic 
  interpretation of expectation values. These two principles together
  are implemented by the requirement that the state\index{state space}
  space is a Hilbert\index{Hilbert space} space, equipped with a
  positive definite inner product. 
\item the local\index{locality}ity (or causal\index{causality}ity)
  principle. This principle expresses the absence of acausal
  influences. It requires the commutativity of quantum observables
  local\index{locality}ized at acausal separation (and is expected 
  to be warranted in the perturbat\index{perturbation theory}ive
  approach if the action functional is a local function of the fields). 
\end{itemize}
In addition, one may (and usually does) require 
\begin{itemize}
\item covariance\index{covariance} under spacetime symmetries (in
  particular, Lorentz invariance of the dynamics), and 
\item stability properties, such as the existence of a
  ground\index{ground state} state (vacuum) or of
  thermal\index{thermal states} equilibrium states.     
\end{itemize}
The critical discussion of these principles themselves (``axioms'') is, 
of course, itself an issue of the axiomatic\index{axiomatic approach}
approaches. For a review, see \cite{BH}. Various axiomatic approaches
(Wightman\index{Wightman theory} QFT, Euclid\index{Euclidean QFT}ean
QFT, Algebraic\index{Algebraic Approach} QFT) may differ in the
technical way the principles are formulated.  Several theorems
establishing (partial) equivalences among these approaches have been
established, such as the Osterwalder-Schrader reconstruction theorem
\cite{OS} stating the precise prerequisites for the invertibility of
the passage from real time QFT to Euclidean QFT (``Wick rotation''),
or the possibility to recover local fields from local algebras \cite{Bos}.

In the Wightman\index{Wightman theory} formulation, one
postulates the existence of fields as operator-valued distributions
defined on a common dense domain within a Hilbert\index{Hilbert space}
space. The field operators should commute at space-like distance and
satisfy a linear transformation law under the adjoint action of a
unitary representation of the Poincar\'e group. Moreover, there should
be a unique Poincar\'e invariant vacuum state which is a
ground\index{ground state} state for the energy operator. The
assumption of local\index{locality} commutativity may be relaxed
admitting anti-commutativity for fermionic fields. One may also relax
the assumption of the vacuum vector, retaining only the positivity of
the energy (unless one is interested in thermal\index{thermal states}
states) in order to describe charged states; in the
algebraic\index{algebraic approach} approach, such theories are most
advantageously regarded as different representations
(superselection\index{superselection sector} sectors) of the same 
field algebra, originally defined in the vacuum representation, see below.
 
Due to the restrictive character of these principles, they typically
are violated in intermediate steps of approximation schemes. One often
has to introduce auxiliary fields without a direct physical meaning as
observables.  

As an illustration, consider the Dirac equation featuring a charged
electron field coupled to the electromagnetic field:
\bea 
\label{dirac}
  i\gamma^\mu(\partial_\mu+ie\, A_\mu)\psi = m\,\psi. 
\eea
The Fermi field $\psi$ satisfies anti-commutation relations and can
therefore not be an observable field strength subject to
causal\index{causality}ity. The vector potential $A_\mu$ is already in
the classical theory not an observable. Related to the gauge
arbitrariness, the vector field cannot be covariantly
quantiz\index{quantization}ed on a Hilbert\index{Hilbert space} space
with a probabil\index{probabilistic interpretation}istic
interpretation. (Other problems related with the promotion to QFT of
classical field products, appearing in evolution equations such as
(\ref{dirac}), will be considered later.) 

The general principles can therefore not be applied to the objects
of basic relations such as (\ref{dirac}). The principles rather apply
to the physical sector of the theory where the typical fields are
current and stress-energy densities or electromagnetic fields, such as 
\bea 
\label{obsfields}
  j^\mu = \bar\psi\gamma^\mu\psi,\qquad T^{\mu\nu} =
  T^{\mu\nu}(\psi,A), \qquad F_{\mu\nu} = \partial_\mu A_\nu
  - \partial_\nu A_\mu. 
\eea
These fields, corresponding to observable quantities, should be
well-defined in a QFT, admitting that the individual quantities on the
right-hand sides of (\ref{obsfields}) turn out to be very ill-defined.

In this spirit, the axiomatic\index{axiomatic approach} approaches
focus directly on the {\em observable} aspects of a theory, that have
an unambiguous and invariant physical meaning, and which should be
computed in order to compare with experiment. They thus strive to
develop analytic strategies to extract these quantities from a given
theory. E.g., the particle\index{particle} spectrum emerges in terms
of poles in renormaliz\index{renormalization}ed correlation functions,
or in terms of the spectrum of the time evolution operator, rather
than as an input in terms of a classical action. The
Haag-Ruelle\index{Haag-Ruelle theory} scattering\index{scattering}
theory showing how the space of scattering state\index{state space}s
(and its structure as a Fock space) is intrinsically encoded, and how
cross section\index{cross section}s are obtained as asymptotic limits
of correlations, was one of the first successes.    

The power of the axiomatic\index{axiomatic approach} approach resides
not least in the ability to derive structural relations among elements
of the theory without the need to actually compute them in a
model. These relations are recognized as necessary consequences of the
axioms. The most familiar examples are the PCT theorem and the
Spin-Statistics theorem, which arise from functional identities among
the Wightman functions due to covariance\index{covariance}, energy
positivity and local\index{locality}ity.  

Another example is the discovery
(``Doplicher-Haag-Roberts\index{Doplicher-Haag-Roberts theory}
theory'') of the coherence among the intrinsic data relating to the
superselection\index{superselection sector} structure (charge
structure). To value this approach, it is important to note that {\em
  if one assumes} (as one usually does) the presence of unobservable
charged fields in a theory, these will typically ``create'' charged
states from the vacuum state $\Omega$. As a specific example,  
\bea 
  \Psi = \psi(f)\,\Omega 
\eea
is an (electrically charged) fermionic state if $\psi(f)$ is an
(electrically charged) Fermi field smeared with some function $f$. 
These states cannot be created by observable fields such as those in
eq.~(\ref{obsfields}), and their charge can be distinguished by 
looking at suitable characteristics of the state functional 
\bea 
  A \mapsto \omega_\Psi(A) \equiv (\Psi\,,\,A\,\Psi) 
\eea 
as the local observables $A$ vary, e.g., when the charge operator $Q$
is approximated by integrals over the charge density $j^0$. 
States of different charge belong to inequivalent representations
(superselection\index{superselection sector} sectors) of the
observables. The DHR\index{Doplicher-Haag-Roberts theory} theory
provides the means to study charged sectors intrinsically, i.e.\
without the assumption of charged fields creating them.

More recently, the DHR\index{Doplicher-Haag-Roberts theory} theory
culminated in the proof (``Doplicher-Roberts reconstruction'') that the
observables along with their charged representations in fact {\em
  determine} an algebra of charged unobservable fields transforming
under a global symmetry group, which create charged sectors from the
vacuum and among which the observables are the invariants under the
symmetry \cite{DR}. Indeed, the presence of Fermi fields, although
these do not correspond to observable quantities, can be inferred (and
their conventional use can be justified) from the existence of
fermionic representations of the bosonic fields of the theory.

At least the relevance of {\em global} symmetry has thus been derived
from the physical principles of QFT. At the same time, the way how
geometric properties of spacetime enter this analysis shows clearly
why the analogous conclusion fails in low
dimension\index{twodimensional QFT}al QFT, opening the way to a much
broader symmetry concept beyond global symmetry groups.  

In realistic models of QFT, the most important symmetry concept is that of 
{\em local} gauge groups, to which we devote a section of its own below. 
Unfortunately, local gauge symmetry is not covered by the
DHR\index{Doplicher-Haag-Roberts theory} theory. 

Axiomatic\index{Axiomatic approach} approaches also allow to
investigate the infrared\index{infrared problem} problem of theories
containing electromagnetism. The infrared problem is due to the fact
that the mathematical description of particle\index{particle} states
as eigenstates of the mass operator  
\bea 
  P_\mu P^\mu\, \Psi = m^2\,\Psi 
\eea
(which is the starting point of the Haag-Ruelle\index{Haag-Ruelle
  theory} scattering theory) 
cannot be used for particles which carry an electric charge. It was
proven under very general conditions \cite{Bu} that electrically
charged sectors contain no eigenstates of the mass operator. Instead
one may use so called particle\index{particle} weights which share
many properties with particle states but are not normalizable \cite{BPS}.  

A more pragmatic way out is the artificial introduction of a photon mass as
a regulator. One computes the cross section\index{cross section}s in
the auxiliary theory and takes the limit of vanishing photon mass for
suitable inclusive cross sections (where ``soft'' photons, i.e.\
photons below an arbitrary small, but finite energy in the final state
are not counted) at the very end. On the conceptual level, this method
involves an exchange of limits. Namely, scattering\index{scattering}
theory in the sense of Haag and Ruelle\index{Haag-Ruelle theory}
amounts to look at distances which are large compared to the   
Compton wavelengths of the particles. The physically relevant limit for 
scattering of electrically charged particles should therefore be to perform 
first the limit for the photon mass and then to go to large distances. 
As was emphasized by Steinmann \cite{St}, it is doubtful whether the
limits may be exchanged.  

The section about
axiomatic\index{axiomatic approach} approaches should not be concluded
without the remark that the complete construction of models fulfilling
all required principles has been achieved with methods described in
Sect.~\ref{construct} below, although presently only in two and three
dimension\index{twodimensional QFT}al spacetime (polynomial
self-interactions of scalar fields, Yukawa interactions with Fermi 
fields). 

Low-dimensional models are of interest as testing grounds for the
algebraic methods and concepts of axiomatic approaches, and to explore
the leeway left by the fundamental principles. Apart from that, since
string theory can in some respect be regarded as (a ten-dimensional
``target space re-interpretation'' of) a conformal quantum field
theory in two dimension\index{twodimensional QFT}s, the exact control
available for a wealth of these models could thus indirectly provide
insight into higher dimensional physics.  

Conformally invariant theories in two dimensions have been
constructed rigorously (and partially classified \cite{KL}) by methods of
operator algebras, especially the theory of finite index subfactors
\cite{LR}. It is here crucial that a ``germ'' of the theory is given,
such as the subtheory of the stress-energy tensor field, and is verified to
share certain algebraic features. Then any local and covariant QFT
which contains this subtheory is strongly constrained, and can be
constructed from certain data associated with the subtheory. Even
if the ``germ'' (as is usually the case) can be realized as a
subtheory of some auxiliary free field theory, the ambient theories
thus constructed cannot be regarded as free theories, i.e., they
``extend the subtheory in a completely different direction''.

Quite recently, a novel scheme for the construction of quantum field
theories has been developped in a genuinely operator
algebraic\index{algebraic approach} approach, which is not based on
quantum fields and some classical counterpart, but on the relation
between the localization of quantum observables and their
interpretation in terms of scattering states. As a consequence of the  
phenomenon of vacuum polarization, this relation is subtle since
interacting local fields can never create pure one-particle
states\index{particle} from the vacuum. The basic new idea stems from
modular theory (see below) by which geometric properties such as
localization in causally independent regions and the action of
Poincar\'e transformations can be coded into ``modular data'' of
suitable algebras. 

Although this is not the place to introduce modular theory \cite{Bo}
to a general audience, we wish to add a rough explanation. There is
a mathematical theorem that the pair of a von Neumann algebra and a
(cyclic and separating) Hilbert space vector determine an associated
group of unitaries and an antiunitary involution, the ``modular data'', 
which have powerful algebraic and spectral properties. In the case of
algebras of covariant quantum observables localized in a wedge region
(any Poincar\'e transform of the region $\vert c\,t\vert <x_1$) and
the vacuum vector, these properties allow to identify the modular data
with a subgroup of the Poincar\'e group and the PCT conjugation. The
joint data for several such wedge algebras generate the unitary
representation of the full Poincar\'e group. Exploiting this algebraic
coding of geometry in the opposite direction, it is in fact possible to
construct a QFT by specifying a distinguished vector in a Hilbert
space and a small number of von Neumann algebras, provided these are
in a suitable ``relative modular position'' to each other to warrant
the necessary relations among their modular data to generate the
Poincar\'e group and ensure local commutativity and energy positivity. 

This opens an entirely new road for the
non-perturbative\index{non-perturbative methods} construction 
of quantum field theory models \cite{SW}. As an example in two
spacetime dimensions, algebras of putative observables localized in
spacetime wedges can be constructed in terms of one-particle
states. Observables with bounded localization are then obtained by
taking intersections of wedge algebras. That this road indeed leads to
the desired construction of interacting theories with a complete
interpretation in terms of asymptotic particle states, has been
established \cite{L} for a large class of models with factorizing
scattering matrices.

\section{The Gauge\index{gauge principle} Principle}\label{gauge}

It happens very often that complicated structures can be more easily 
accessed by introducing redundant quantities. The extraction of the
relevant information then requires a notion of equivalence. In
fundamental physics it is the notion of a local\index{locality}
interaction which forces the introduction of redundant structures. To
ensure that the observable quantities do not influence each other at a
distance, one wants to describe their dynamics by field equations
which involve only quantities at the same point. But it turns out that
this is possible only by introducing auxiliary quantities, such as
gauge potentials in electrodynamics. This difficulty already exists in
classical field theory, and it complicates considerably the structure
of classical general relativity.   

Classical gauge theories describe the interaction of gauge fields 
(understood as connections of some principal bundle) and matter fields 
(described as sections in associated vector bundles). The interaction
is formulated in terms of covariant derivative\index{covariant
  derivative}s and curvatures. (In this way, the rather marginal gauge
symmetry of Maxwell's electrodynamics is turned into a paradigmatic
symmetry principle determining the structure of interactions.) The
combination 
\bea 
  D_\mu=\partial_\mu+ie A_\mu
\eea 
providing the coupling between the fields in (\ref{dirac}) is a
covariant derivative which ensures that the equation is invariant
under the gauge transformation     
\bea 
\label{gauge transformation}
  \psi(x)&\mapsto& \hbox{e}^{ie\alpha(x)}\,\psi(x) \nonumber \\ 
  A_\mu(x) &\mapsto& A_\mu(x) - \partial_\mu\alpha(x),
\eea 
that is, $D_\mu\psi$ transforms in the same way as $\psi$ itself. The
field strength tensor $F_{\mu\nu}$ in (\ref{obsfields}) is obtained
through the commutator of two covariant derivative\index{covariant
  derivative}s, i.e.\ geometrically speaking, the curvature.  

The presence of this group of automorphisms (\ref{gauge transformation}) 
of the bundle (gauge transformations) makes the description redundant,
and only the space of orbits under the automorphism group corresponds
to the relevant information.

In quantum field theory, the very concept of gauge theories is
strictly speaking not well defined, because of the singular character
of pointlike localized  quantities. These singularities are absent in
the lattice\index{lattice approximation} approximation (see
Sect.~\ref{construct}). There matter fields are attached to the
vertices, and gauge fields are as parallel transporters attached to
the links. 
 
In perturbat\index{perturbation theory}ion theory (see
Sect.~\ref{perturb}) additional auxiliary structure has to be invoked
in order to be able to use the canonical formalism. Namely, the Cauchy
problem in gauge theories is not well posed because of the ambiguities
associated with time dependent gauge transformations. Therefore one
has to introduce a gauge fixing term in the Lagrangean which makes the
Cauchy problem well posed, and so called ``ghost\index{ghost fields}
and antighost'' fields which interact with the gauge field in such a
way that the classical theory is equivalent to the original gauge
theory. This auxiliary theory is quantiz\index{quantization}ed on a
``kinematical Hilbert\index{Hilbert space} space'' ${\mathcal H}$ which is not
positive definite. The observables of the theory are then defined as
the cohomology of the BRST\index{BRST method} transformation $s$ which
is an infinitesimal symmetry of the theory with $s^2=0$ (see, e.g.,
\cite{Wein}). More precisely, $s$ is implemented as a graded
derivation by a charge operator $q$ such that $q^2=0$, the observables
are those local operators that commute with $q$: 
\bea 
  q \, A = A \, q, 
\eea
physical state\index{state space}s are those annihilated by it:
\bea 
  q\,\Psi_{\rm phys} = 0,
\eea
and two physical states are equivalent if they differ by a state in
the image of it: 
\bea 
  \Psi_1 - \Psi_2 \;\in\; q\,{\mathcal H}.
\eea
The BRST\index{BRST method} method ensures that the equivalence
classes of physical states form a positive-definite
Hilbert\index{Hilbert space} space 
\bea {\mathcal H}_{\rm phys} = \hbox{Ker }q\,\big/\,\hbox{Im }q, \eea
and the observables are well-defined operators on ${\mathcal H}_{\rm phys}$.

We will see in Sect.~\ref{perturb} that within perturbation theory,
BRST\index{BRST method} gauge theories are distinguished by their good
behaviour under renormaliz\index{renormalization}ation.

It is not clear how the gauge\index{gauge principle} principle should enter 
the axiomatic\index{axiomatic approach} formulations. These approaches
focus on the observables of a quantum system, while gauge fields are
{\em per se} unobservable. Put differently, one should ask the
question which observable features tell us that a QFT is a gauge
theory. In the abelian case, there is of course the characteristic
long-range nature of Gauss' law, but there is no obvious equivalent in
the nonabelian case, i.e.\ when $\psi$ in (\ref{gauge transformation})
is replaced by a multiplet and the phase factor
$\hbox{e}^{ie\alpha(x)}$ by a unitary matrix. Could there be, in
principle, an alternative description of, say, QCD\index{QCD} without
gauge symmetry?  

There are of course experimental hints towards the color symmetry, 
ranging from particle\index{particle} spectroscopy over total cross
section\index{cross section} enhancement factors to ``jets'' in
high-energy scattering\index{scattering}. In algebraic\index{algebraic
  approach} QFT, the counterpart of these observations is the analysis 
of the global charge structure of a theory, i.e.\ the structure of the
space of state\index{state space}s. 

The DHR\index{Doplicher-Haag-Roberts theory} theory of
superselection\index{superselection sector} sectors is precisely an
analysis of the charge structure entirely in terms of the algebra of
observables. As we have seen, it leads to the derivation of a symmetry
principle from the fundamental principles of QFT (see
Sect.~\ref{axiom}), but the result pertains to global symmetries
only. The case of local gauge symmetries is still open. Yet, a local
gauge theory without confinement\index{confinement} should possess
charged states in nontrivial representations of the gauge group. If
the theory has confinement, but is asymptotic\index{asymptotic
  freedom}ally free, then its gauge group should become visible
through the charge structure of an appropriate short-distance limit of
the observables \cite{BH}. It is therefore expected that gauge
symmetry, if it is present, is not an artefact of the perturbative
description but an intrinsic property coded in
algebraic\index{algebraic approach} relations among observables. 

\section{The Field Concept}\label{field}

It is the irony of Quantum Field Theory that the very notion of a
``quantum field'' is not at all obvious. The field concept has been
developped in classical physics as a means to replace the
``action\index{action at a distance} at a distance'' by perfectly
local\index{locality} interactions, mediated by the propagating
field. Classical fields, such as the electromagnetic fields, can be
observed and measured locally. On the other hand, in quantum field
theory one usually interprets measurements in terms of
particle\index{particle}s. The fields used in the theory for the
prediction of counting rates, appear as (very useful, undoubtedly)
theoretical constructs, imported from the classical theory. But what
is their actual status in reality?   

The conventional particle\index{particle} interpretation requires that
a given state behaves like a multi-particle state at asymptotic
times. A closer look shows that this feature may be expected only in
certain circumstances, say, in a translationally invariant theory in
states close to the vacuum. Once one leaves these situations, neither
the concept of a vacuum (ground\index{ground state} state of the
energy) nor that of particles (eigenstates of the mass operator) keep
a distinguished meaning, as may be exemplified by the occurrence of
Hawking\index{Hawking radiation} radiation, by the difficulties of a
notion of particles in thermal\index{thermal states} states, and last
not least, in the infrared\index{infrared problem} problem. 

The field concept, on the other hand, keeps perfect sense in all 
known cases. Fields may be understood, generally speaking, as means to 
interpret quantum theoretical objects in geometrical terms. In 
Minkowski space, they may assume the form of distributions whose 
values are affiliated to the algebras of local observables and which 
transform covariantly under Poincar\'{e} transformations.  Here, the
test function $f$ plays the role of a ``tag'' which keeps track of the 
local\index{locality}ization of the associated field operator $\varphi(f)$. 
In a generally covariant\index{general covariance} framework (see
Subsect.\ \ref{CST}), fields can be viewed abstractly as natural
transformations from the geometrically defined functor which associates
test function spaces to spacetime manifolds, to the functor which
associates to every spacetime its algebra of local\index{locality}
observables \cite{BFV}.    

On the mathematical side, the field concept leads to hard problems in
the quantum theory. They are due to the quantum fluctuations of
localized observables which diverge\index{ultraviolet singularities}
in the limit of pointlike localization. But in
perturbat\index{perturbation theory}ion theory as well as in
algebraic\index{algebraic approach} quantum field theory one has
learned to deal with these problems, the most difficult aspect 
being the replacement of ill-defined pointwise products by the
operator product expansion.  

In free field theory on Minkowski space, one associates to every 
particle\index{particle} a field which satisfies the field equation.
While in this case, the use of the term ``particle'' for the
associated field is perfectly adequate, the analogous practice for
fields which appear in the classical equation of motion of 
{\em interacting} field theory is justified only in special cases. 
It may happen (this seems to be the case in asymptotic\index{asymptotic
  freedom}ally free theories) that in a short distance limit, the
analogy to the particle--field correspondence of free field theory
becomes meaningful. In theories which become free in the infrared
limit, a similar phenomenon happens at large distances; then the
scattering\index{scattering} data can be directly interpreted in terms
of these distinguished fields. 

In general, however, besides the observable fields one uses a whole
zoo of auxiliary fields which serve as a tool for formulating the
theory as a quantiz\index{quantization}ation of a classical Lagrangean
field theory. Such a formulation may not always exist nor must it be
unique. In the functional\index{functional methods} (``path
integral'') approach to QFT, such auxiliary fields (which are not
coupled to external sources) may be regarded as mere integration
variables. The most powerful functional techniques involve deliberate
changes in such variables (introduction of ``ghost\index{ghost fields} 
fields'', BRST\index{BRST method} transformations, or the
renormaliz\index{renormalization}ation program by successive
integration over different energy scales). While this is by far the
most successful way to construct models, at least in the sense of
perturbat\index{perturbation theory}ion theory, the intrinsic physical
significance of these auxiliary fields is unclear, and it would be
misleading to think of them in terms of particle\index{particle}s in a
similar way as discussed before.   

The delicacy of the field concept in Quantum Theory, contrasted with the
clarity of the classical field concept, may be just one aspect of the more
fundamental question: Is a quantum theory necessarily the
quantiz\index{quantization}ation of a classical theory? Does it always
have a classical limit (think of QCD\index{QCD}, for the sake of
definiteness), and can it be reconstructed from its classical limit? 

\section{The Perturbative Approach to QFT}\label{perturb}

The main approximative schemes for relativistic QFT are
Perturbat\index{Perturbation Theory}ion Theory (or other expansions
like the $1/N$ approximation), and lattice\index{lattice
  approximation} approximations of Euclid\index{Euclidean QFT}ean
functional\index{functional methods} integrals. All these
approximations of QFT are based on the idea of
``quantiz\index{quantization}ation of a classical field
theory''. Perturbation theory proceeds by producing a formal power
series expansion in a coupling  constant, hoped to be asymptotic to a
QFT yet to be constructed, and therefore requires  weak couplings;
lattice\index{lattice approximation} approaches can in principle also
treat strongly coupled regimes, using cluster expansions or Monte Carlo
simulations; although numerical simulations of lattice QFT are limited 
to rather coarse lattices, aspects of the continuum and infinite volume
limits can be studied. As far as comparisons are possible, there seems
to be little doubt about the basic consistency among different approaches.

Our discussion in this section will mainly pertain to Perturbation Theory. 
This is a general scheme a priori applicable to any Lagrangean with a 
``free'' and an ``interaction'' part. Characteristic limitations to the
scheme arise, however, through various sources: 

First of all, there is the need to
``renormaliz\index{renormalization}e'' the single terms of the
perturbat\index{perturbation theory}ive expansion. This is the
procedure to fix the parameters of the theory to their physical
values, thereby also avoiding any infinities that occur if one
proceeds in the traditional way using ``bare'' parameters. One must
demand that renormalization can be achieved without the introduction
of infinitely many new parameters which would jeopardize the
predictive power of the theory. This necessity restricts the
admissible form of the interaction Lagrangean. Provided the polynomial 
order in the fields is limited (depending on the spacetime dimension,
and on the spins of the fields involved), a simple ``power\index{power
  counting} counting'' argument (controlling the behaviour of
potentially divergent terms in terms of the momentum dependence of
propagators and interactions) ensures
renormaliz\index{renormalization}ability. For spins larger than 1,
there are no interactions in four dimensions which are renormalizable
by power counting. (This fact also prevents the direct incorporation
of gravit\index{quantum gravity}ational fields into the  
perturbat\index{perturbation theory}ive scheme.) In the presence of
additional symmetries which ensure systematic cancellations of
divergent terms, renormaliz\index{renormalization}ability might be
shown in spite of a failure of the power\index{power counting}
counting criterium (but in supersymmetric\index{supersymmetry}
perturbative gravity the desired effect seems to fail beyond the first
few lowest orders).   

For the theory of elementary particles, experiments have revealed the 
prime relevance of vector (spin 1) couplings, starting with parity
violation in the weak interaction which could be explained by $V-A$
but not by scalar and tensor couplings. The idea that vector couplings
are mediated by vector fields lies at the basis of the Standard
Model. For interactions involving massless vector fields, however,
there is a conflict between local\index{locality}ity,
covariance\index{covariance}, and Hilbert\index{Hilbert space} space
positivity, while massive vector fields do not possess couplings which
are renormaliz\index{renormalization}able by power\index{power counting}
counting. This is due to the fact that the necessary decoupling of 
modes which otherwise would give rise to states of negative norm,
changes the large-momentum behaviour of the propagator. 

The only successful way to incorporate vector fields into a
perturbat\index{perturbation theory}ive QFT is to treat them as gauge
fields, with couplings which are necessarily gauge couplings (see
Sect.~\ref{gauge}). Thus, the gauge\index{gauge principle} principle  
imposes itself through the inherent limitations of the perturbative
scheme \cite{Sto}. However, it brings about several new problems which
have to be solved in turn: the unphysical degrees of freedom can be
eliminated by cohomological methods (``BRST\index{BRST method}
theory'', see Sect.~\ref{gauge}) which at the same time can be used to  
systematically control the preservation of gauge invariance. While
gauge invariance forbids the introduction of explicit mass terms for
the vector fields, masses can be generated by coupling to a
Higgs\index{Higgs mechanism} field with
``spontaneous\index{spontaneous symmetry breaking} symmetry
breakdown''. That this can indeed be done in a way which keeps the
theory renormaliz\index{renormalization}able in spite of the bad
power\index{power counting} counting behaviour of massive propagators,
is one of the great achievements of the perturbat\index{perturbation
  theory}ive Standard Model.   

In the process of renormaliz\index{renormalization}ation there may
appear ``anomal\index{anomalies}ies'' which break symmetries present
in the classical theory. While anomalies per se are not problematic
(and may even be phenomenologically desirable), anomalies of the {\em
  gauge} symmetry will spoil the renormalizability. Their absence has
therefore to be imposed as a consistency condition. In chiral gauge
theories, it can be achieved by a suitable choice of representations
of the gauge group (particle multiplets).  

The circumstance that the ``cascade of problems'' outlined in the
preceding paragraph can in fact be consistently overcome within the
setting of perturbat\index{perturbation theory}ive QFT, and in
excellent agreement with the phenomenology of High Energy Physics,
gives enormous confidence in the basic correctness of the
approach. The Standard Model precisely exhausts the leeway admitted by
the perturbative approach. 

Besides the renormaliz\index{renormalization}ation problems caused by
ultraviolet\index{ultraviolet singularities} singularities,
perturbative QFT has infrared\index{infrared problem} problems, when
the free theory used as the starting point contains massless
particle\index{particle}s. In Quantum Electrodynamics
(QED\index{QED}), the infrared problem can be traced to the
computational use of particles with sharp masses which is illegitimate
in the presence of massless particles (see Sect.~\ref{perturb}). 

A more severe kind of infrared\index{infrared problem} problem arises
in theories like QCD\index{QCD}; here it is due to the fact that the
fields (quarks and massless gluons) do not correspond to the massive
particle\index{particle}s (hadrons) presumably described by the full,
non-perturbative theory. A fully consistent solution of these
problems, i.e.\ the confinement\index{confinement} of hadronic
constituents, can therefore not be expected in a perturbative
treatment. If the confinement problem can be addressed at all, then
only by non-perturb\index{non-perturbative methods}ative methods (see
the next section). However, effects like the deviations from naive
scaling of hadronic structure functions have been successfully predicted by
perturbative methods. 

The infrared\index{infrared problem} problems of
perturbat\index{perturbation theory}ion theory may be circumvented by
the use of interactions which are switched off outside some compact
region of spacetime. This leads to the concept of causal
perturbat\index{causal perturbation theory}ion theory which was
developed by Epstein and Glaser \cite{EG} on the basis of previous
ideas of St\"uckelberg and Bogoliubov. This approach is crucial for a
consistent treatment of quantum field theory on curved\index{curved
  spacetime} spacetimes. On Minkowski space it allows a perturbative 
construction of the algebra of observables. The infrared problem then
is the physical question on the {\em states} of the theory, such as the
existence of a ground\index{ground state} state, the
particle\index{particle} spectrum, thermal\index{thermal states}
states and so on.  

Whether one considers the rationale for the gauge\index{gauge
  principle} principle in the Standard model outlined above (see also
Sect.~\ref{gauge}) to be logically cogent, depends on the implicit
expectations one imposes on the formal structure of a QFT. In any
case, the Standard Model is by no means uniquely determined by these
constraints. QED\index{QED} and QCD\index{QCD} are completely
self-consistent subtheories (that is on the level of a formal
perturbative expansion); the subtheory of electro-weak\index{weak
  interaction} interactions is consistent provided the gauge
anomal\index{anomalies}ies are eliminated by suitable charged
multiplets. The gauge groups themselves may be considered as free
parameters of a model, as long as anomaly cancellation is possible. 

The possibility of Grand\index{grand unification} Unification and/or
Supersymmetric\index{supersymmetry} Extensions is an aesthetic feature
of the Standard Model, for which, however, there is no fundamental
physical need, nor is it required for reasons of mathematical
consistency. QFT alone presumably cannot answer the question why there
are so many ``accidental'' free parameters (notably the mass matrices,
or Yukawa couplings, according to the point of view) in the theory of
fundamental interactions.  

To conclude this section, we should point out that, as far as model
building is concerned, the limitation to
renormaliz\index{renormalization}able interactions might be too
narrow. There are perturbat\index{perturbation theory}ively
non-renormalizable\index{non-renormalizable theories} model theories   
in which nontrivial fixed points have been established, meaning that the
theories are non-perturb\index{non-perturbative methods}atively
renormalizable \cite{GK}.  

\section{The Constructive Approach to QFT}\label{construct}

In spite of its tremendous numerical success, the
perturbat\index{perturbation theory}ive scheme to evaluate QFT
approximately suffers from a severe defect: it provides answers  
only in the form of formal, most likely {\it divergent} power series.
The usual answer to this is that the series is an asymptotic expansion. 
But aside from the problem where to truncate the series in order to 
convert the formal power series into numbers, there is the fundamental 
question: asymptotic to what? There are well-known cases (such as the 
so-called $\Phi^4_4$ theory of a self-interacting scalar field $\Phi$
in four spacetime dimensions) in which the perturbation expansion, 
according to the accumulated knowledge, is {\it not} an asymptotic 
expansion to any QFT and it may very well be that the most successful of 
all QFT's, Quantum Electrodynamics\index{QED}, also suffers from this disease.
  
The axiomatic\index{axiomatic approach} approach, on the other hand,
does not answer the question whether the axioms are not empty, i.e.\
whether any {\it nontrivial} QFT's satisfy them.

The constructive\index{constructive QFT} approach is in principle
addressing both of these problems. On the one hand it attempts to show
that the axiomatic framework of QFT is not empty, by mathematically
constructing concrete nontrivial examples satisfying these axioms, and
on the other hand it provides non-perturb\index{non-perturbative
  methods}ative approximation schemes that are intimately related to
the attempted mathematical constructions; the prime example are the
lattice\index{lattice approximation} approximations to QFT's. Even
where the goal of a mathematical construction of  models satisfying
all the axioms is not (yet) attained, this kind of approximative
scheme differs in a fundamental way from the formal perturbative
expansions: it produces approximate numbers which, if all goes right,
{\it converge} to a limit that would be the desired construction.

The constructive\index{constructive QFT} approach (see for instance
\cite {GJ}) is based on a modification and generalization of Feynman's
``sum over histories''. The main modification is the transition form
the indefinite Lorentz metric of Minkowski spacetime to a
Euclid\index{Euclidean QFT}ean metric; the return to the physical  
Lorentzian metric is expected to be manageable via the so-called 
Osterwalder-Schrader reconstruction \cite{OS} (see Sect.~\ref{axiom}). 
The approach starts from a classical field theory, with dynamics 
specified by a Lagrangean. Formally one then procedes by writing an 
ill-defined functional\index{functional methods} integral over all
field configurations, weighted with a density given in terms of the
classical action $S=\int {\mathcal L} \,dx$ depending on some fields
collectively denoted by $\Phi$; the expectation value of an ``observable''
${\mathcal O}[\Phi]$ (a suitable function of the fields) would be given by
\bea
  \langle {\mathcal O}\rangle =\frac{1}{Z}\int {\mathcal D}\Phi\;
  {\mathcal O}[\Phi]\; e^{-S[\Phi]}\ .
\eea
Here the symbol ${\cal D}\Phi$ is supposed to indicate a (nonexisting) 
Lebesgue measure over the set of all field configurations $\Phi$ and $Z$ a 
normalization constant.

To make mathematical sense of this, the theory first has to be 
``regularized'' by introducing a finite spacetime volume and deleting or 
suppressing high frequencies (by an ``ultraviolet\index{ultraviolet
  singularities} cutoff''). The job of the constructive field theorist
then consists of controlling the two limits of infinite volume
(``thermodynamic limit'') and restoring the high frequencies
(``ultraviolet limit'') by removing the cutoff; the latter can only be
done if the parameters of the Lagrangean (and the observables of the
theory) are made cutoff dependent in a suitable way -- this procedure
is the non-perturb\index{non-perturbative methods}ative version of
renormaliz\index{renormalization}ation. 

The constructive\index{constructive QFT} program has been completed
only in spacetime dimension\index{twodimensional QFT}s less than four,
but at least in these unrealistic cases it has shown that axiom
systems such as Wightman\index{Wightman theory}'s are not vacuous
for interacting theories. In these low-dimensional cases it has also
given a justification to the perturbat\index{perturbation theory}ive
expansion by showing that it produces indeed an asymptotic expansion
to the constructed QFT's. 

A particularly useful way of introducing an
ultraviolet\index{ultraviolet singularities} cutoff consists in
replacing the spacetime continuum by a discrete structure, a
lattice\index{lattice approximation}. Together with the introduction
of a finite spacetime volume one thereby reduces quantum field theory
to a finite dimensional ``integral'' (the quotation marks indicate
that this ``integral'' is just some linear form for the fermionic
degrees of freedom). In other words, QFT has been reduced to
quadratures. The advantage of this is that QFT thereby becomes  
amenable to numerical evaluation; there is a whole industry of lattice 
field theory exploiting this fact, most notably in approximately 
evaluating the theory of strong\index{strong interaction}
interactions, QCD\index{QCD}.  

But the lattice\index{lattice approximation} approach is very
important also for more fundamental reasons: it is the only known
constructive\index{constructive QFT} approach to a
non-perturb\index{non-perturbative methods}ative definition of gauge
field theories, which are the basis of the Standard Model. The
constructive approach and the numerical procedures to extract infinite
volume and continuum information from finite lattices are closely parallel:

Typically a lattice\index{lattice approximation} model produces its
own dynamically generated scale $\xi$ (``correlation length'') which,
unlike the lattice spacing, has a physical meaning. It may be defined
-- after the thermodynamic limit has been taken -- by the exponential
decay rate of suitable correlation functions, such as 
\bea
\label{defxi}
  \xi=-\lim_{n\to\infty} \frac{1}{|n|}\ln \langle\, \Phi(0)\Phi(n)\,\rangle\, ,
\eea
where $\Phi(\cdot)$ stands for a field of the lattice theory
and $n$ is a tupel of integers labeling lattice points.

In a finite volume version, finite volume effects disappear exponentially
fast, like $\exp(-L/\xi)$, with the size $L$ of the volume. The
thermodynamic limit can then be controlled numerically and often also
mathematically, borrowing techniques from classical Statistical Mechanics.

The next step is to identify the dimensionless number $\xi$ with a
physical standard of length (e.g., some appropriate Compton wave
length, say 1 fm), such that $\xi$ lattice\index{lattice
  approximation} spacings equal 1 fm. The lattice points can then be
relabeled by $x_i=(n_i/\xi)$ fm where the coordinates $x_i$ have now
acquired the dimension of length. Taking the lattice spacing to zero
(i.e.\ taking the continuum limit) then amounts to sending the
correlation length to infinity while keeping $x_i$ fixed. 
The $n$-point correlation functions of a field in the continuum 
should therefore be defined as limits of the form
\bea 
  \langle\, \varphi(x_1)\ldots \varphi(x_n)\,\rangle = 
  \lim_{\xi\to\infty} \langle\, \Phi([x_1]\xi) \ldots \Phi([x_n]\xi)\,\rangle 
  \;Z(\xi)^{-\frac{n}{2}}\ 
\eea
where $x=[x]$ fm, and $\varphi(x)$ is the resulting continuum quantum field. 

So the continuum limit requires to drive the parameters of the system
(such as the coupling constants) to a point of divergent correlation
length, i.e.\ a critical point in the language of Statistical
Mechanics. $Z(\xi)$ is a ``field strength
renormaliz\index{renormalization}ation'' needed to prevent the limit
from being 0.  

This procedure makes it clear that the lattice\index{lattice
  approximation} spacing is a derived dynamical quantity proportional
to $1/\xi$, not something to be specified beforehand. The inverse of
the correlation length in the chosen physical units is the mass gap of
the theory in physical units. The procedure of choosing the
dynamically generated scale as the standard of length or mass leads
generally to a phenomenon usually attributed to special features of
perturbation theory: ``dimensional transmutation''. Let us explain
this in a simple case, QCD\index{QCD} with massless quarks: the only
parameter of the lattice theory is the gauge coupling; since we find
the continuum limit at the (presumably unique) critical point, this
ceases to be an adjustable parameter. Instead we obtain a free scale
parameter by the freedom of choosing a certain multiple of the  
correlation length as the standard of length (or a certain multiple of 
the inverse correlation length as the standard of mass). So we have 
traded a dimensionless parameter (the coupling constant) for a parameter 
with dimensions of a mass (e.g., the mass of the lightest particle).

Quite generally the particle\index{particle} spectrum of any QFT is
extracted by looking at exponential decay rates of suitable
correlation functions; when applied to QCD\index{QCD} the
lattice\index{lattice approximation} approach has been reasonably
successful in reproducing the observed spectrum of baryons and mesons. 
It has also been successfully extended to the computation of
weak\index{weak interaction} decay matrix elements of hadrons. All
this gives us confidence that QCD is indeed an appropriate description
of the strong\index{strong interaction} interactions. 

On the constructive\index{constructive QFT} side, the success with
gauge theories in four dimensions has been much more modest, even
though some impressive mathematical work towards control of the
continuum limit has been done by Balaban \cite{Bal}. 

\section{Effective quantum field theories}\label{effect}

In applications one often encounters the term
``effective\index{effective theory} field theory''.
We can distinguish three different meanings:

(1) the result of an exact Renormaliz\index{renormalization}ation
Group (RG) transformation applied to a quantum field theory in the
sense discussed before, 

(2) an approximate quantum field theory that is supposed to give a 
good approximation to a certain assumed QFT,

(3) a phenomenological theory that is not to be taken seriously beyond a 
certain energy; in this case it does not matter if the theory arises from 
a bona fide QFT by some approximation or by integrating out high momentum 
modes.

The notion (1) is at least conceptually very clear. The idea is to
start with an already constructed well-defined quantum field theory
and then to apply an exact ``Renormaliz\index{renormalization}ation
Group step''. This means that one performs the part of the
functional\index{functional methods} integral (which has been made
well-defined before) corresponding to the ``hard'' (i.e.\ high
momentum, fast varying) part of the fields, formally 
\bea
  \exp(-S_{\rm eff}[\Phi_{\rm soft}])= 
  \frac{1}{Z} \int {\mathcal D}\Phi_{\rm hard}\;
  \exp(-S_{\rm eff}[\Phi_{\rm soft}+\Phi_{\rm hard}])\ .
\label{rg}
\eea
and also performs some rescalings of fields and spacetime variables.
The combination of the integration in (\ref{rg}) and this rescaling 
constitutes one Renormalization Group step.
The resulting ``effective\index{effective theory} theory'' describes
exactly the same physics as the original full theory when applied to
the soft (low momentum, slowly varying) degrees of freedom. It is
clear that this may be ``effective'', but it is not efficient because
it requires control of the full theory before one can even start.

Of course the RG step sketched above can be iterated; thereby one 
generates the semigroup usually called
Renormaliz\index{renormalization}ation Group. 

A more useful variation of the RG idea is used in
Constructive\index{Constructive QFT} Quantum Field Theory (see for
instance \cite{Bal,GK}).  Here one starts with a regularized version
of the theory, defined with a high momentum cutoff; one then performs
a number of RG steps as indicated above until one reaches a predefined
``physical scale'' leading to an effective\index{effective theory} low
energy theory still depending on the cutoff. In the final step one
attempts to show that the low energy effective theory has a limit as
the cutoff is removed; this requires adjusting the parameters of the
starting ``bare action'' such that the effect of the increasing number
of successive Renormaliz\index{renormalization}ation Group steps is
essentially compensated.

The notion (2) is widely used to describe the low energy physics of
QCD\index{QCD} (assumed to exist as a well-defined Quantum Field
Theory even though this has not been shown so far). Specific examples are: 
\begin{itemize}
\item
``Effective Chiral Theory'' \cite{Leu} to describe the interactions of 
the light pseudoscalar mesons,
\item
``Heavy Quark Effective Theory'' (HQET) \cite{Ei}, in which the effect 
of the heavy (charmed, bottom and top) quarks is treated by expanding 
around the limit where their masses are infinite, 
\item 
``Nonrelativistic QCD'' (NRQCD) \cite{Cas,Bod} used in particular 
for bound state problems of heavy quarks.  
\end{itemize}
For an overview over various applications of these ideas see \cite{Pi}.
\vskip5mm
Examples for notion (3) are the old Fermi theory of weak\index{weak
  interaction} interactions (before the electro-weak part of the
Standard Model was known). A more modern example is presumably the
Standard Model itself, because it contains the scalar self-interacting
Higgs\index{Higgs mechanism} field which suffers from the presumed
triviality of  $\Phi^4_4$ theories; the same applies to any other
model involving Higgs fields. One often finds the words ``something  
is only an effective\index{effective theory} theory''; this expresses
the fact that the author(s) do not want to claim that their model
corresponds to a true quantum field theory.

\section{Gravity}\label{gravity}

Given the state of affairs for the Standard Model of elementary
particles, being comfortably well described by Quantum Field Theory as 
outlined in the previous sections, the ``missing link'' in our present
conception of fundamental physics is the incorporation of the
gravit\index{quantum gravity}ational interaction into quantum physics
(or vice versa).  

For a review of classical Gravity, we refer to the contribution by Ehlers
to this volume. Empirically, Gravity is a theory valid at macroscopic
scales only, and it is well known that, if extrapolated to very small
scales (the Planck\index{Planck length} length), it becomes
substantially incompatible with the quantum
uncertainty\index{uncertainty} principle (``quantum energy
fluctuations forming virtual black\index{black holes} holes''). This
suggests that at small scales Gravity needs modification, although one
might as well argue conversely, that at small scales Gravity modifies
Quantum Theory (by acting as a physical regulator for the
UV\index{ultraviolet singularities} problems of QFT, or possibly in a
much more fundamental manner). The truth is not known, and one might
expect that neither Quantum Theory nor Gravity will ``survive''
unaffected in the ultimate theory.  

Empirical evidence for this case is of course extremely poor due to the
smallness of the Planck\index{Planck length} length. The most
promising candidate for empirical evidence about effects of Quantum
Gravity are astronomical observations of matter falling into
supermassive black\index{black holes} holes, or cosmological remnants
of the very early universe. On the theoretical side, it is generally
expected that black hole physics (Hawking\index{Hawking radiation}
radiation and Bekenstein entropy) represents the crucial point of
contact. It appears very encouraging that both major approaches
(String Theory and Canonical Quantum Gravity, see below), in spite of
their great diversity, make more or less the same predictions on this
issue. But it should be kept in mind that also Hawking radiation of
black holes is far from being experimentally accessible.  

The attempt to incorporate the gravit\index{quantum gravity}ational
interaction into Quantum Theory raises severe conceptual
difficulties. Classical gravity being a field theory, QFT is expected
to be the proper framework; but QFT takes for granted some fixed
background\index{background} spacetime determining the
causal\index{causality} structure, as one of its very foundations,
while spacetime should be a dynamical agent in gravity theory. This
argument alone does not preclude the logical possibility of
perturbat\index{perturbation theory}ive
quantiz\index{quantization}ation of gravity around a fixed background,
but on the other hand, the failure of all attempts so far which split
the metric into a classical background part and a dynamical quantum
part (cf.\ Sect.~\ref{axiom}), should not be considered as a complete
surprise, or as a testimony against QFT.  

On the other hand, the existing arguments against the
quantiz\index{quantization}ation of gravity within a conventional QFT
framework are not entirely conclusive. They are based on the most
simple notion of renormaliz\index{renormalization}ability which
demands that the renormalization flow closes within a finite space of
{\em polynomial} couplings, thus giving rise to the limitation by
power\index{power counting} counting. It is conceivable, and there are
indications that something in this way actually occurs \cite{Reu},
that a renormalization flow closes within a finite space of suitable 
{\em non-polynomial} Lagrangeans (which are present in classical
Einstein gravity anyway). In this case, the renormalized theory also
would contain only finitely many free parameters, and would have the
same predictive power as a theory with polynomial interactions.

Taking the geometrical meaning of gravitational fields seriously, it
is clear that the framework of QFT has to be substantially enlarged in
order to accomodate a quantum theory of Gravity. It is
an open question whether this can be done by formal analogies between 
diffeomorphism invariance and gauge symmetry. 

\subsection{QFT on gravitational background spacetime}\label{CST}

An intermediate step on the way towards a theory of Quantum Gravity is a
semiclassical treatment, where ``matter'' quantum fields are defined
on classical curved\index{curved spacetime} spacetimes. This situation
brings along severe technical and conceptual problems, since crucial
tools of quantum field theory in flat spacetime (energy-momentum
conservation, Fourier transformation and analyticity, Wick rotation,
particle\index{particle} interpretation of asymptotic
scattering\index{scattering} states) are no longer available due to
the lack of spacetime symmetries.    

Considerable progress in this direction has been made notably
concerning the problem of the absence of a distinguished
ground\index{ground state} state (the vacuum). In globally hyperbolic
spacetimes, the ground state can be substituted by a class of
state\index{state space}s (Hadamard states) which guarantee 
the same stability properties of quantum fields, and allow for a
similar general set-up of causal perturbat\index{causal perturbation
  theory}ion theory as in flat space \cite{HW}. Of crucial importance
is the incorporation of the principle of general
covariance\index{general covariance}. It is realized as a covariant
functor which associates to every globally hyperbolic spacetime its
algebra of observables and which maps isometric embeddings of
spacetimes to homomorphic embeddings of algebras. The interpretation
of the theory is done in terms of covariant fields, which are
mathematically defined as natural transformations from a geometrically
defined functor which associates to every spacetime its test function
space to the functor describing the quantum field theory \cite{BFV}. 

One may include into the set of quantum fields also the fluctuations
of the metric field. One then has to impose the consistency condition
that the result does not depend on the chosen split of the metric into
a background\index{background} field and a fluctuation field (this is
essentially Einstein's equation in quantum field theory). One may hope
to obtain in this way reliable information on the ``back reaction'' of
the energy of the quantum matter on the background. It remains,
however, the bad power\index{power counting} counting behaviour of
quantum gravity which might point to limitations of the
perturbat\index{perturbation theory}ive approach.    

\subsection{Non-commutative spacetime}

Taking into account the expectation that local\index{locality}ization
should be an operational concept which at very small scales is limited
by the interference between quantum and gravit\index{gravity}ational
effects, models of non-commutative\index{non-commutative spacetime}
spacetimes have been formulated which exhibit an intrinsic
local\index{locality}ization uncertainty\index{uncertainty}.  
While these are definitely not more than crude models, in which gravity 
is not itself present but just motivates the localization uncertainty,
it could be established that they are compatible with Quantum Field
Theory; contrary to widespread hopes, however, the quantum structure of 
spacetime does not act as a ``physical regulator'' at the
Planck\index{Planck length} scale for the
ultraviolet\index{ultraviolet singularities} behaviour of quantum
field theory \cite{BDFP}

\subsection{Canonical Quantum Gravity}

Other approaches to Quantum Gravity focus on the purely 
gravit\index{quantum gravity}ational self-interaction. The most
prominent ones, going under the name ``Canonical quantum gravity'',
are built upon the geometric nature of classical gravity. In these
approaches, the dynamics of three-dimensional (space-like) geometries
is studied in a canonical framework. However, due to general
covariance\index{general covariance}, the dynamics turns out to be
strongly constrained, giving rise to severe complications. (See the
contribution of Giulini and Kiefer to this volume.)  

Within the general framework of canonical approaches, Loop Quantum
Gravity (LQG) has been pursued and developped furthest as a model for the
structure of quantum spacetime \cite{A} (see also the contributions of
Thiemann and of Nicolai and Peeters to this volume). It is asserted
that the model can be supplemented by any kind of ``conventional''
matter (e.g., the Standard Model). It therefore denies every ambition
towards a unified or unifying theory.  

For these reasons, critical questions confronting the model with
the requirements for a ``true'' theory of Quantum Gravity, are more
or less void. As for its intrinsic consistency and mathematical
control, the model meets rather high standards, consolidating and
improving previous attempts of canonical
quantiz\index{quantization}ation of Gravit\index{quantum gravity}y.

The model predicts that geometric observables such as areas and
volumes are quantized, with lowest eigenvalue spacings of the order
of the Planck\index{Planck length} size. This feature appears most
promising in that quantum deviations from classical geometry are
derived as an output, with no classical
(``background\index{background}'') geometry being used as an input. 

On the other hand, one of the most serious flaws of LQG is the lack of
understanding of its relation to gravity ``as we know it'', i.e.\ the
construction of semiclassical state\index{state space}s in which
Einstein's General Relativity at large scales is (at least in some
asymptotic sense) restored. 

Another, presumably related drawback of LQG (like any other model
within the canonical approach to Quantum Gravity) is that in the
physical Hilbert\index{Hilbert space} space, once it has been
constructed, the Hamiltonian vanishes. Thus, the question of the
nature of ``time'' evolution of the quantum gravitational states
is presently poorly understood.  

\subsection{String Theory}

A detailed discussion of successes and problems of String Theory
will be given in contributions of Louis, Mohaupt and Theisen to
this volume. 
We will here restrain ourselves to some questions focussing on the
intrinsic structure and the conceptual foundations of String Theory, 
which appear quite natural to ask having in mind the benefits of
axiomatic\index{axiomatic approach} approaches in the case of
QFT. Even if some of our questions might appear immodest, the theory
being still under construction,  they should be settled in some sense
before String Theory can be considered as a mature theory. 

String Theory is a quantum theory naturally including
gravit\index{quantum gravity}ational degrees of freedom in a unified
manner along with ``conventional'' matter. Gravitons and other
particles arise as different ``zero modes'' of strings which are the
fundamental objects; higher vibrational modes would correspond to
undetected heavy particles (with masses far beyond accelerator energies). 
This fact is the prominent source of enthusiasm with the theory. (For
a critical comparison of the achievements of String Theory and of Loop
Quantum Gravity as candidates for the quantum theory of gravitation,
see e.g., \cite{Smo}.) 

The theory can successfully reproduce scattering\index{scattering}
cross section\index{cross section}s for gravitons as they are expected
in the lowest orders of perturbat\index{perturbation theory}ion 
theory with the Einstein-Hilbert action. In contrast to perturbation
theory (cf.\ Sect.~\ref{perturb}), the theory is believed to have 
a better UV\index{ultraviolet singularities} behaviour due to the
finite size of the string, but its alleged finiteness (or
renormaliz\index{renormalization}ability) could not be established 
with the increasing understanding of higher order contributions to
string theory.   

On the phenomenological side, it was hoped that a unified theory
including the Standard Model of elementary particles would naturally
emerge as an effective\index{effective theory} theory at low (compared
to the Planck\index{Planck length} scale) energies, but these hopes
were considerably reduced by an enormous number of possible ``string
vacua'', destroying the predictive power of the theory.

String theory was originally formulated in a
perturbat\index{perturbation theory}ive scheme, where spacetime
appears just as a classical background\index{background}. The 
dynamics of the string moving in this background is given by a
two-dimension\index{twodimensional QFT}al conformal quantum field 
theory (organizing its internal degrees of freedom), whose consistency
requires the background to satisfy Einstein's equations. In the course
of time it became clear, that a consistent formulation of string
theory has to take into account non-perturbative structures like
duality symmetries, including the need to introduce higher-dimensional
objects (``branes''). The presence of these classical objects is
expected to be related to the question (although still far from
answering it) of the quantum nature of spacetime itself \cite{Hor}.    

Non-perturb\index{non-perturbative methods}ative formulations of
string theory are in the focus of most modern developments. Yet, the
mathematical structure of non-perturbative string theory and the
picture of spacetime and quantum gravit\index{quantum gravity}y which
emerges, are at the present time not yet well understood beyond a huge
body of heuristic imagination, based on the various duality symmetries
of string theory and the ``holographic principle\index{holographic
  principle}'' concerning the quantum degrees of freedom of general 
relativity. A most fascinating recent development is Maldacena's
conjecture which states that non-perturbative string theory could be
``equivalent'' (in a sense involving duality) to a quantum field
theory, possibly even in four dimensions. The theory which started off
to supersede QFT may in the end be equivalent to a QFT!  

As a computational scheme, String Theory is highly constrained and
determined by its internal consistency. For this reason, it is often
claimed to be a ``unique'' theory, hence it makes little sense to
``axiomatize'' String Theory in a similar way as Quantum Field Theory
was axiomatized (Sect.~\ref{axiom}). Nevertheless, the justification of its
computational rules deserves some critical scrutiny. 

The central question is: Which are the fundamental insights into the
nature of physical laws (principles) that are implemented by String
Theory? Is String Theory unique in doing so, or is it possibly only
{\em one} consistent realization of the same principles?  
Accepted principles such as Quantum Uncertainty\index{uncertainty},
Local\index{locality}ity and General Relativity should be transcended
by the new principles without recourse to (classical) notions outside
the Theory. 

An important ``message'' from algebraic\index{algebraic approach} QFT
is that the intrinsic invariant structure of the quantum observables
are their algebraic relations such as local\index{locality}
commutativity, rather than their description in terms of
fields. (Neither the classical action nor Feynman diagrams are
intrinsic; field equations and canonical commutation relations cannot
even be maintained after quantiz\index{quantization}ation.) 
The ``concrete'' (Hilbert space) representations of these ``abstract''
algebraic relations determine the physical spectrum (masses, charges).   

In this spirit, one would like to identify the intrinsic elements
of String Theory, and the structural relations which hold {\em a priori} 
among them. An intrinsic characterization would also turn claims such
as the Maldacena conjecture into predictions that can be verified (or
falsified).

It is generally agreed that a classical background\index{background}
manifold should not appear in an ultimate formulation of String
Theory. This is not only because the metric is expected to fluctuate,
so that it is impossible to describe its expectation values in a
particular state by a classical geometry. Since spacetime structures
smaller than the string size cannot be probed, and hence cannot have
an operational meaning, String Theory is expected to produce a
radically new concept of spacetime. 

While String Theory is an S-matrix theory, that is, in a suitable
limit it admits the computation of ``on-shell'' particle
scattering\index{scattering} amplitudes, ``off-shell'' String Field
Theory has been rigorously constructed only without interactions
\cite{Dim}. The resulting theory may be viewed as a collection of
infinitely many ``ordinary'' quantum fields, but their
local\index{locality} commutativity cannot be ensured in a covariant 
way. The reason is that the constraints on the string degrees of
freedom prevent the construction of sharply or only compactly
localized observables on the physical (positive-definite)
Hilbert\index{Hilbert space} space out of string fields defined on an
indefinite space. (In view of the previous remark, this conflict with
the classical spacetime concept should not  come as a surprise.) With
interactions, the description in terms of an infinite tower of quantum
fields is expected to survive, but the structure of the interactions
(string corrections to the effective action) goes beyond the framework of
local Lagrangean quantum field theory. Correspondingly, String Field
Theory (even in a regime where gravity can be neglected) is not
expected to be a QFT in the sense of Sects.~\ref{axiom} or
\ref{perturb}.  

On the other hand, String Theory exhibits a new fundamental symmetry
called ``duality''. The Maldacena conjecture suggests that under a
duality transformation, String Theory could turn into a quantum field
theory. A clarification of the precise non-perturbative meaning of
this conjecture is highly desirable, not least in view of the numerous
and far-reaching implications drawn from it.

As an example, $T$-duality, relating vibrational and winding modes of a
string, is a most characteristic symmetry of String Theory. With the
help of $T$-duality one can understand how a string fails to be able
to probe certain singularities of a classical
background\index{background} \cite{Hor}. Positing duality symmetry as
an abstract fundamental symmetry is a promising candidate for an
intrinsic structure of the theory which can be formulated without
recourse to the classical picture of a string embedded into spacetime.  

As for the intrinsic texture of String Theory (assuming it to be a
consistent theory), it would be desirable to understand in which sense 
its subtheories (``spacetime without matter'', ``QFT without Planck
scale gravity'') are separately consistent, or rather only
effective\index{effective theory} theories obtained by a singular
limit, which is regulated by the full theory.  

While some of these questions might indeed rather reflect the authors'
personal rooting in QFT (and also some lack of understanding of String
Theory), we think that they are urgent enough that expert string
theorists should provide answers in order to legitimate String Theory
as a candidate for the Fundamental Unified Theory of all interactions. 

\section{Conclusions and Outlook}

Whether the remaining gaps in the theory are merely of technical
nature, or rather signal a fundamental shortcoming of Quantum Field
Theory, is not known at present, and is by many researchers not
considered as the most urgent question. 

Instead, the prime concern at present is the clash between Gravity 
and Quantum Theory, whose unification is considered as the (last)
``missing link'' in our conception of fundamental physics.
There are promising candidate theories to achieve this ambitious goal,
but none of them shares the same conceptual clarity as has been
attained for Quantum Field Theory, nor are there empirical data
available favouring or disfavouring either of them. 

Unlike almost every historical precedent, the guiding principle at the 
frontiers of research in fundamental physics is therefore mainly
intrinsic consistency, rather than empirical evidence. Every active
researcher should be aware of the delicacy of such a situation.

It should be remarked that, while various lines of research presently
pursued call basic notions such as Geometry and Symmetry into
question, the basic rules of Quantum Theory are never challenged.  
One may be tempted to ascribe this fact to the solidity of our
conceptual understanding of Quantum Physics, developped over several 
decades not least in the form of Quantum Field Theory.

\vskip15mm\newpage
{\large\bf Note to the references} 
\vskip3mm
There is a long list of standard textbooks on Quantum Field Theory. The
subsequent list of references leaves out most of them, as well as much
of the ``classical'' research articles. Instead, it includes a number
of less well-known articles, stressing some points which are relevant in our
discussion but which do not belong to the common knowledge about
quantum field theory.

\end{document}